\documentclass[fleqn,usenatbib]{mnras}

\usepackage{newtxtext,newtxmath}
\usepackage[T1]{fontenc}

\DeclareRobustCommand{\VAN}[3]{#2}
\let\VANthebibliography\thebibliography
\def\thebibliography{\DeclareRobustCommand{\VAN}[3]{##3}\VANthebibliography}

\usepackage{graphicx}	% Including figure files
\usepackage{amsmath}	% Advanced maths commands
\newcommand{\Msun}{\,\ensuremath{\mathrm{M}_\odot}}
\usepackage{xcolor}
\title[White Dwarf Luminosity Function]{The imprint of the first stars on the faint end of the white dwarf luminosity function}

\author[Dzięcioł, Hartwig \& Yoshida]{
Bartosz Dzięcioł,$^{1}$\thanks{E-mail: bartosz.dzieciol@phys.s.u-tokyo.ac.jp}
Tilman Hartwig,$^{1,2,3}$
 Naoki Yoshida,$^{1,3,4}$
\\
$^{1}$Department of Physics, School of Science, The University of Tokyo, Bunkyo, Tokyo, 113-0033, Japan \\
$^{2}$Institute for Physics of Intelligence, School of Science, The University of Tokyo, Bunkyo, Tokyo, 113-0033, Japan \\
$^{3}$Kavli Institute for the Physics and Mathematics of the Universe (WPI), The University of Tokyo Institutes for Advanced Study,\\
The University of Tokyo, Kashiwa, Chiba, 277-8583, Japan\\
$^{4}$Research Center for the Early Universe, School of Science, The University of Tokyo,
Bynkyo, Tokyo, 113-0033, Japan\\
}

\date{Accepted XXX. Received YYY; in original form ZZZ}

\pubyear{2023}

\begin{document}
\label{firstpage}
\pagerange{\pageref{firstpage}--\pageref{lastpage}}
\maketitle

% Abstract of the paper
\begin{abstract}
Population III stars are characterized by extremely low metallicities as they are thought to be formed from a pristine gas in the early Universe. Although the existence of Population III stars is widely accepted, the lack of direct observational evidence hampers the study of the nature of the putative stars. In this article, we explore the possibilities of constraining the nature of the oldest stars by using the luminosity function of their remnants -- white dwarfs. We study the formation and evolution of white dwarf populations by following star formation in a Milky Way-like galaxy using the semi-analytic model \textsc{a-sloth}. We derive the white dwarf luminosity function by applying a linear Initial-Final Mass Relation and Mestel's cooling model. The obtained luminosity function is generally in agreement with available observations and theoretical predictions -- with an exponential increase to a maximum of $M_{\mathrm{abs}} = 16$ and a sudden drop for $M_{\mathrm{abs}} > 16$. We explore the uncertainties of our model and compare them to the observational estimates. We adopt two different models of the initial mass function of Population III stars to show that the faint end of the luminosity function imprints the signature of Population III remnants. If the feature is detected in future observations, it would provide a clue to Population III stars and would also be an indirect evidence of low- to itermediate-mass Population III stars. We discuss the challenges and prospects for detecting the signatures. 
\end{abstract}

\begin{keywords}
white dwarfs -- stars: luminosity function, mass function -- stars: population III -- stars: evolution -- stars: formation -- methods: numerical
\end{keywords}

%%%%%%%%%%%%%%%%%%%%%%%%%%%%%%%%%%%%%%%%%%%%%%%%%%

%%%%%%%%%%%%%%%%% BODY OF PAPER %%%%%%%%%%%%%%%%%%

\section{Introduction}
\label{sec:introduction}

White dwarfs (WDs) are the remnants of cores of low to intermediate-mass stars -- they can be formed from stars of masses between approximately 0.4 and 10\Msun~depending on other conditions, including progenitor star metallicity. They are extremely dense, composing of an electron-degenerate matter. The white dwarf luminosity function, which we refer to as WDLF or simply LF, is a population distribution of white dwarfs with respect to their luminosity. In general, 
it characterises the differential counts of luminosity (or absolute magnitude), often divided by the total volume in which the considered WDs exist. 
Various studies have been conducted to derive WDLF both from observations \citep{Leggett_1998, Liebert_2005, Harris_2006} and theoretical models, especially by Monte Carlo simulations \citep{Winget1987-ln, Wood_09, Berro_99, Torres_02, Garc_a_Berro_2004, Torres_2012}. In some cases, the disk and halo WDs are distinguished, and separate LFs are computed for both populations. Most of the previous studies agree on the general shape of the function -- with the counts increasing exponentially from brightest WDs ($M_{\mathrm{abs}} \approx$ 5--6) to the maximum at $M_{\mathrm{abs}} \approx$ 15--16 and then sharply dropping. The published observational data is still very limited in the faint end, which makes it especially worth further investigation. It is expected that upcoming releases of GAIA data \citep{GAIAHR} as well as data from future surveys will greatly 
help our understanding on WDLF \citep{Garc_a_Berro_2016}.

%IMF (maybe shift below):
The WDLF is shaped by two main ingredients: the initial mass function (IMF) of stars and their stellar evolution. While the IMF of metal-enriched stars in the MW 
has been studied in detail, virtually nothing is known about the IMF of Pop~III stars. Based on the non-detection of any surviving metal-free stars in the MW, the lower mass limit of the Pop~III IMF is derived to be $\gtrsim 0.65$ \Msun~with 95\% confidence \citep{hartwig15,ishiyama16,magg18,rossi21}. 
There are no direct observational constraints on the formation of massive Pop~III stars. The chemical compositions of several extremely metal-poor stars suggest that the first stars exploded as core-collapse supernovae with up to 100\Msun~\citep{ishigaki18,hartwig23}. However, the shape of the Pop~III IMF is largely unconstrained. We thus resort to examining two different Pop~III IMFs in the present study: one fiducial IMF with a logarithmically flat slope between 5--260\Msun~based on \citet{hartwig22,uysal23}, and another bottom-heavy case with a Salpeter slope in the mass range 0.65--260\Msun.

IFMR (Initial-Final Mass Relation) characterises the dependence of the mass of the remnant WD on the progenitor mass. Although the exact shape of IFMR is still largely uncertain, it is generally an increasing function, i.e. heavier progenitor stars produce heavier WDs, and moreover, for heavier WDs the percentage of mass loss is higher. IFMR can be obtained in either semi-empirical or purely theoretical way. Some former models argue for a simple linear IFMR, at least for the low mass progenitors \citep{Kalirai_2008, Catalán20081693, K_lebi_2013}. Recent studies show that IFMR is not only dependent on the progenitor star metallicity in a non-monotonic way, with the minimum of percentage of mass loss for metallicity around $Z = 0.04$ \citep{Meng08, Romero15, Zhao2012-lc}, but also non-linear and non-monotonic \citep{Ferrario_2005, Cummings2015-ru, Marigo_2017, Cummings2018, Marigo_2020, Marigo22}, possibly with more than one non-monotonic kink \citep{Marigo22a}. IFMR for Pop III stars is a widely unexplored topic, however, there has been recent attempts to obtaining it theoretically \citep{lawlor13}.

The earliest important model of WD cooling mechanism was developed by Mestel and published as \citet{Mestel1952}, then developed by him and other authors in 1960's \citep{Mestel1967-le}. Contemporary models take into account various aspects which Mestel was unaware of, and generally describe the cooling process by four stages -- neutrino cooling, fluid cooling, crystallisation and Debye cooling. The processes were studied in detail by various authors especially for carbon oxygen WDs, while cooling of helium WDs and oxygen-neon WDs attracted significantly less attention. The cooling of carbon-oxygen WDs is described in \citet{Isern_2013}. Numerical simulations were also performed to predict cooling curves of WDs of different physical characteristics \citep{Salaris_1997}. Although the Mestel's model assumptions and results differ from contemporary understanding of WD cooling mechanism, up to this day his results are surprisingly accurate given his model's simplicity. Because the Mestel's cooling curve can be easily expressed in analytical form, it is still a valuable tool to investigating WDs. 

However it is clear that a significant part of the white dwarf population reside in binary systems, the exact share of white dwarfs in such systems is uncertain. The overall binary fraction between 25\% and 40\% is found in the literature, but this value also depends on factors such as mass of white dwarf and belonging to specific region of the galaxy (halo or disk) \citep{holberg09, toonen17, cukanovaite23}. In any case within this range the influence of binary interactions on WDLF might be significant. Binarity of WD population can have a few different effects on WDLF. Firstly, the time between the birth of a star and formation of a WD might differ from the normal evolutionary models in such systems because of matter exchange. Secondly, a certain part of WDs in binary systems can be He white dwarfs which undergo cooling which is not described by Mestel's model. Lastly, supernovae explosions lead to complete eradication of some part of white dwarf population. In this paper, we assume all stars and product WDs to be born and evolve single and we calculate the WDLF without considering the binarity of the WD population, while remembering that it could have some effect on the results. 

The aim of this work is to try a different approach to producing a WDLF by employing a recently published tool, \textsc{a-sloth} \citep{Magg_2022} to perform a computer simulation of star formation in Milky Way-like galaxy (from now on we will refer to Milky Way as MW). The tracked WDs should be divided into two groups in accordance to their progenitor star population -- let us call them Pop~II and Pop~III WDs respectively. As \textsc{a-sloth} does not distinguish Pop I from Pop~II stars, some of the stars from the first category will also be considered Pop~II stars here. We try to examine whether any imprint of Pop~III WDs could be potentially found in the WDLF. This will be the first approach to follow the evolution of Pop~III WDs in detail and to estimate their contribution to the WDLF.

The potential imprint of Pop~III stars on the WDLF is small. In this paper, we want to investigate the systematic uncertainties in modelling the WDLF in order to guide future investigations into the faint end of the WDLF.

\section{Methods}

\subsection{A-Sloth}
\label{subsec:sloth}
We simulated the formation of stars in the MW with the semi-analytical model \textsc{a-sloth} (Ancient Stars and Local Observables by Tracing Halos, \citealt{Magg_2022})\footnote{\url{https://gitlab.com/thartwig/asloth}}. The model is ideally suited for our purpose since it traces individual populations of stars and is calibrated based on six independent observables \citep{hartwig22,uysal23}. 

\textsc{a-sloth} simulates the formation of stars on top of dark matter merger trees. We explore 30 different realizations of MW-like galaxies from the Caterpillar Project \citep{Griffen_2016} and compare the results. These merger trees offer sufficient resolution in mass and time to resolve the formation of the first stars in minihalos at high redshift. Moreover, the variance of these merger trees allows us to investigate the effect of cosmic variance on the WDLF.

We ran the simulations assuming two different Initial Mass Functions for Pop~III stars -- the fiducial one, with a slope of $-1.0$ and minimum mass of 5\Msun~and the bottom-heavy one, with a Salpeter IMF (slope $-2.3$) and minimum mass of 0.65\Msun. For Pop~II stars, Salpeter IMF was assumed in both cases.

For every realization and every time-step, we tracked the numbers of stars being formed. We used the stellar lifetimes as a function of ZAMS mass from \citet{schaerer02} for metal-free stars and from \citet{stahler05} for metal-enriched stars to calculate the time of transition to the remnant.

We assume that all stars with masses between 0.5\Msun~and 9.75\Msun~produce carbon-oxygen WD remnants. While it is a simplification because some of the heaviest WDs might be oxygen-neon WDs instead and some of the stars of masses between 8--9.75\Msun~might even produce a neutron star instead of a WD, it should not affect the model very significantly -- carbon-oxygen WDs still make up a vast majority of remnants \citep{10.3389/fspas.2022.815517}. It is unlikely that stars with masses heavier than 9.75\Msun~produce white dwarfs \citep{doherty14}. We investigated the impact of chemical composition of the carbon-oxygen WD's core on the results by repeating calculations for extreme values. Following this assumption, the number of WDs commencing at each time-step was collected, with regard to the progenitor star population, mass and metallicity. We then further processed the data to create LFs. Summing up the data for all the time-steps gives the total number of WDs expected in each realization. The average total number of WDs in all realizations is $(1.82 \pm 0.70) \times 10^{10}$ for both fiducial IMF and bottom-heavy IMF. The average total number of Pop~III WDs is $(2.3 \pm 0.9) \times 10^4$  for the fiducial IMF and $(6.2 \pm 2.3) \times 10^6$ for the bottom-heavy IMF. This means that the average share of Pop~III WDs is $1.2$ ppm for fiducial IMF and $340$ ppm for bottom-heavy IMF, however for different realizations it ranges between $0.8$--$1.8$ ppm and $218$--$489$ ppm, respectively. The complete data is available upon request.

The total final stellar mass of the galaxy in all the thirty realizations varies significantly from $2.34 \times 10^{10}$\Msun~to $9.85 \times 10^{10}$\Msun~for both fiducial and extreme IMF. The average is $(5.67 \pm 1.87) \times 10^{10}$\Msun. We compare this value to the previous MW evolution model from \citet{Boissier_1999}. That model gives a value of $3.8 \times 10^{10}$\Msun, which is reasonably close to our average. As we consider 30 different realizations of a MW-like galaxy, each of them predicting different total final stellar mass, the discrepancies can be observed. 

\subsection{Obtaining White Dwarf luminosity function}
\label{subsec:lumfuntheor}

\subsubsection{White Dwarf cooling function}
\label{subsubsec:wdcoolingfun}

For transforming the times of WD formation into their luminosities, we used simple Mestel's cooling model \citep{Mestel1952, Mestel1967-le}. The luminosity of a WD of a given age was assumed to follow the equation
\begin{equation}
    L(t) = L_{\odot} \frac{M}{\Msun} \left( \frac{10^8}{A (t/\textrm{yr})}\right)^\frac{7}{5}.
	\label{eq:mestel}
\end{equation}
In equation~(\ref{eq:mestel}) $M$ is the mass of a WD, $A$ is the mean atomic weight of its core and $t$ is the WD age. While we simply assumed that for the carbon-oxygen WDs $A=14$, WD progenitors were divided into nineteen equally sized mass bins and the WD masses were then obtained by employing the linear IFMR given by
\begin{equation}
    M_{\mathrm{WD}}=0.109M_{0} + 0.394,
	\label{eq:ifmr}
\end{equation}
where $M_{0}$ is mass of a progenitor and $M_{\mathrm{WD}}$ is mass of the corresponding WD. This equation is a linear fit presented in \citet{Kalirai_2008}. The more advanced, progenitor metallicity-dependent model was found not to change the results in any significant way and thus given up in favor of this simpler, linear model. Comparison of Mestel's model with more advanced MESA model is presented in Section~\ref{subsec:mesamestel}.

\subsubsection{Data processing}
\label{subsubsec:processing}

%\begin{equation}
%    M_{\mathrm{bol}} = -2.5 \log (L) + 71.1974.
%	\label{eq:lummag}
%\end{equation}

We converted the obtained luminosities to absolute magnitudes and calculated WDLF as a function $\textrm{d} \log{N} / \textrm{d} M_{\mathrm{abs}} (M_{\mathrm{abs}})$.
In the last steps of data processing, we approximated linearly the data for different progenitor masses with the magnitude resolution of 0.001, added up and further smoothed linearly at the bright end to avoid non-physical remnants -- 'steps' coming from adding up datasets with different bright-end limits. The exact shape of this part of the plot is approximate and therefore should be treated with reserve. However, it contains data about white dwarf counts that sum up to the total number of white dwarfs in our model and as such it was kept. On all figures it is presented with dotted lines. See Fig.~\ref{fig:m_comparison_ext} in Appendix for the plot presenting all the mass bins separately. Analogically, a sample LF presenting products of progenitors of different metallicities separately (bottom-heavy Pop~III IMF case) is presented in Fig.~\ref{fig:z_comparison_ext} in the Appendix. 

\section{Results}
\label{sec:results}

\subsection{White dwarf luminosity function}
\label{subsec:lumfun}

\begin{figure}
	\includegraphics[width=\columnwidth]{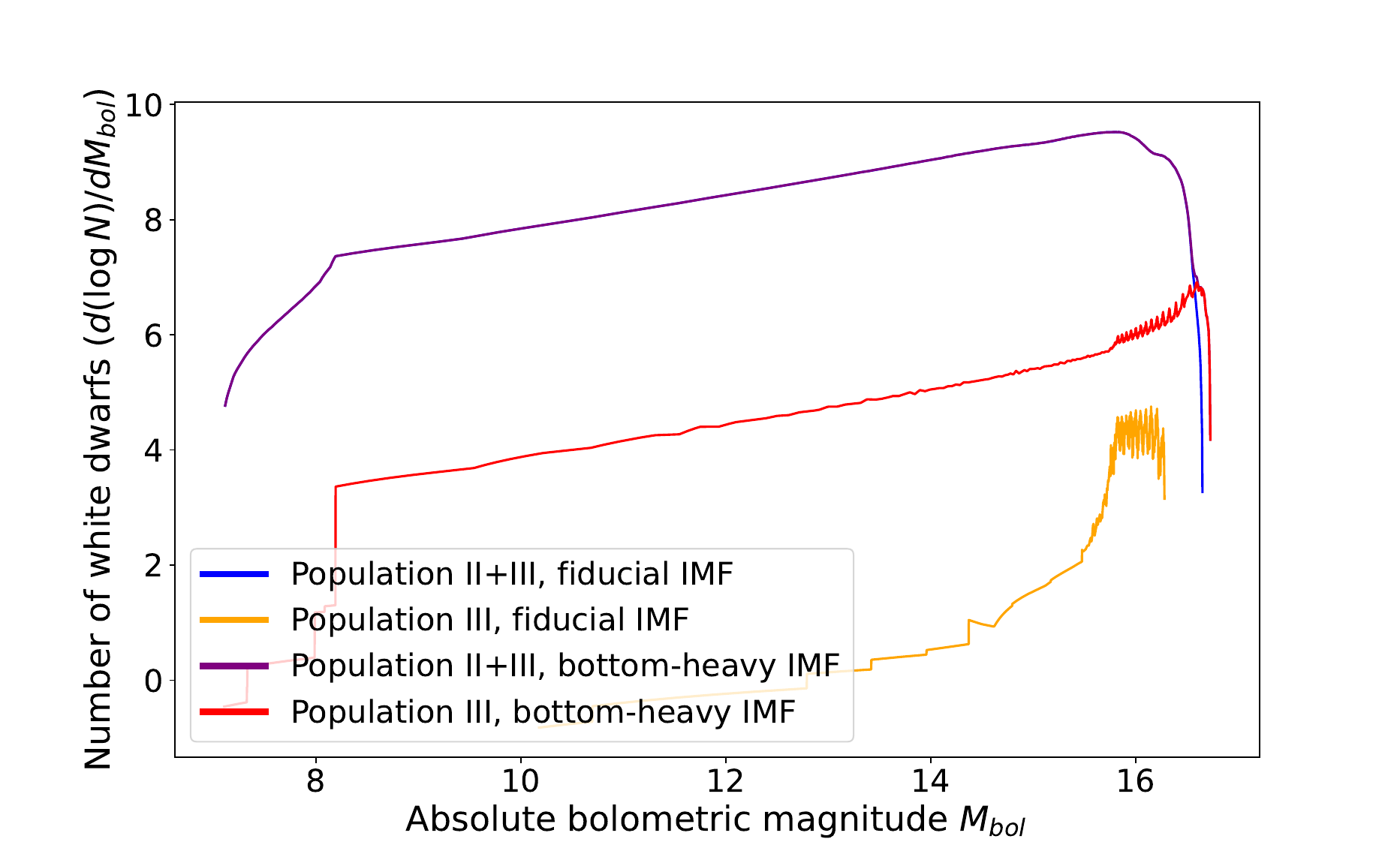}
    \caption{Obtained WDLF divided with respect to progenitor star population and Pop~III IMF. Only one MW realisation was chosen to ensure the clarity. The blue and purple lines represent the data for all white dwarfs, respectively for fiducial and bottom-heavy IMF cases while the yellow and red lines include only data for WDs descending from Pop~III stars respectively for fiducial and bottom-heavy IMF case. By looking at the figure it becomes clear that while for the fiducial IMF case the influence of Pop~III star-descending WDs on the overall IMFR is little, for the bottom-heavy stars it is important, especially in the faint end.}
    \label{fig:II_III_all}
\end{figure}

The obtained WDLFs for thirty different MW-like galaxy realizations for fiducial and bottom-heavy IMF are presented in Fig.~\ref{fig:II_III_all}. The overall shape of the LFs is the same in both cases -- for $M_{\mathrm{abs}} > 8$ the function is increasing in an approximately linear way to reach the maximum at $M_{\mathrm{abs}} \approx 8$. In the faint end, it rapidly drops -- this drop is limited by the age of first WDs in the Galaxy. We can see that in the bright end, for $M_{\mathrm{abs}} \approx$ 7--8 mag the slope becomes steeper -- it is in fact expected to behave this way, but the rapid descent is an effect of mass binning and necessary linear approximations of function in this range (see Section~\ref{subsubsec:processing}). 

\begin{figure}
	\includegraphics[width=\columnwidth]{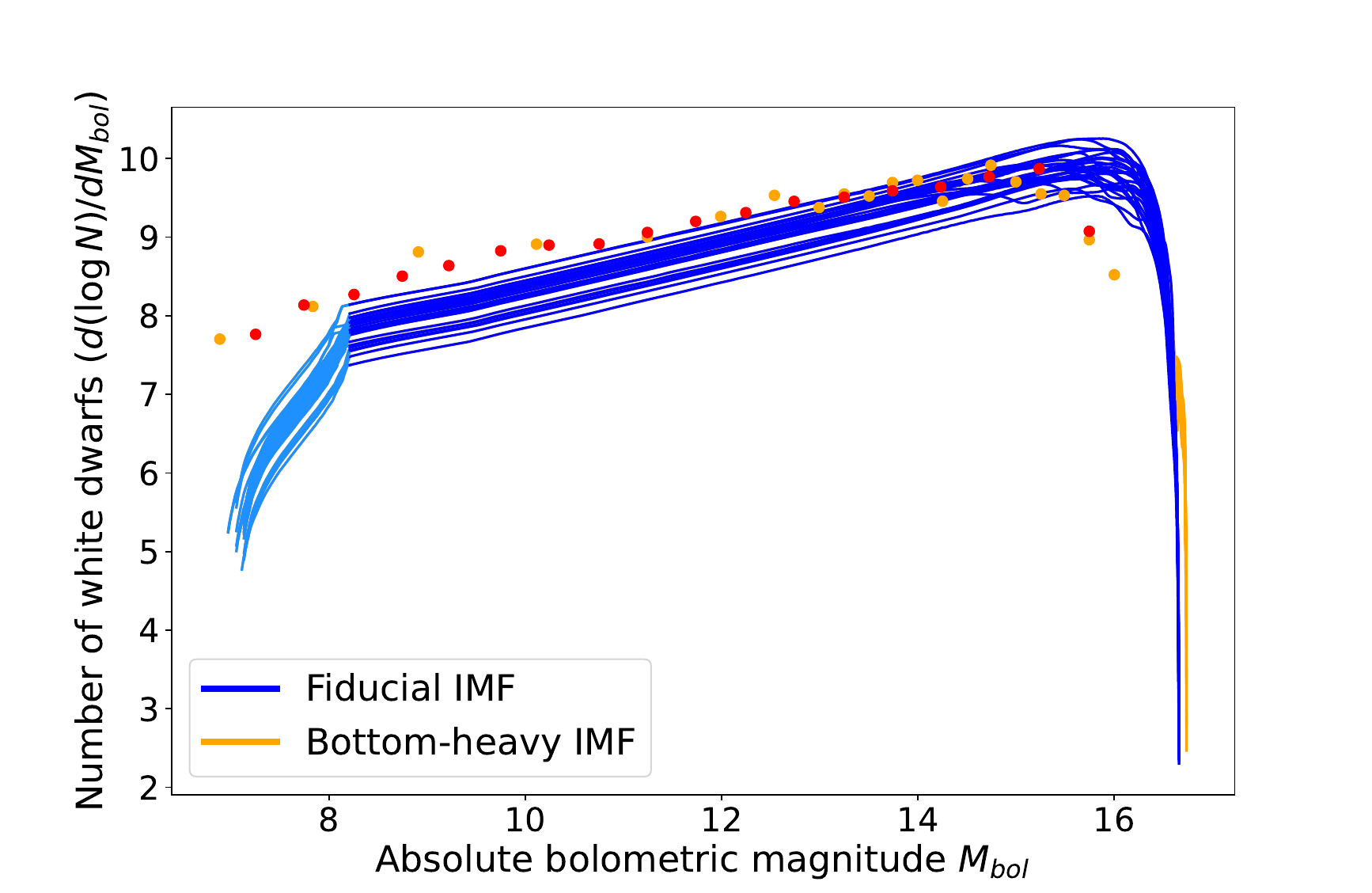}
    \caption{LFs of dwarfs coming from Pop~II and III progenitor stars. Different blue lines represent different MW realisations for fiducial IMF while different orange lines represent different MW realisations for bottom-heavy IMF. The part of the plot presented in light blue is a linear approximation which influenced its shape}
    \label{fig:IIffe_III_all}
\end{figure}

\begin{figure}
	\includegraphics[width=\columnwidth]{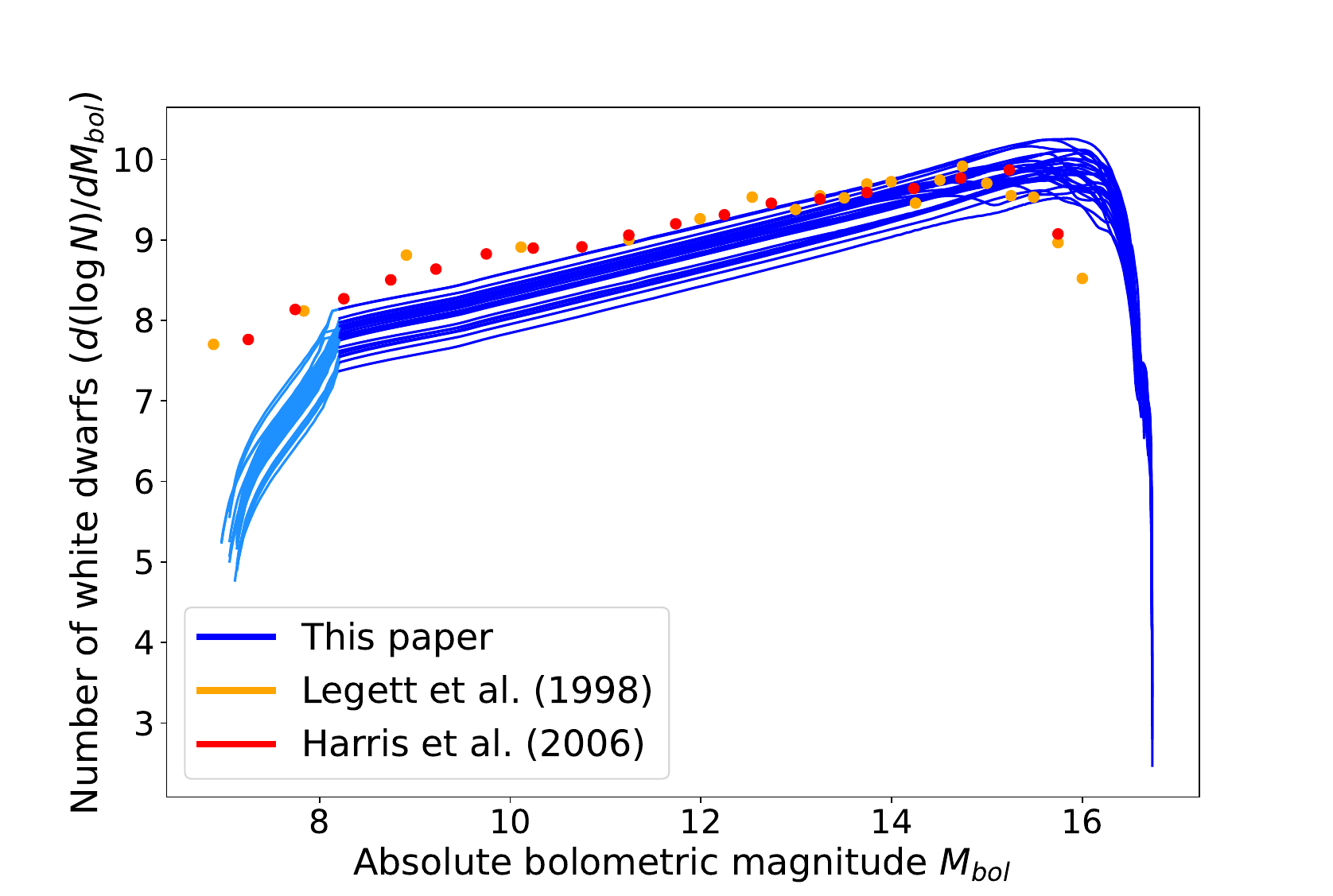}
    \caption{Comparison of obtained WDLF with observational results. The blue lines represent the LFs from this work. Yellow dots make the WDLF from \citet{Leggett_1998} while red dots represent the data from \citet{Harris_2006}. The observational data was normalized to the total number of WDs predicted in this work. The part of the plot presented in light blue is a linear approximation which influenced its shape.}
    \label{fig:obscomp}
\end{figure}

Fig.~\ref{fig:obscomp} presents the bottom-heavy IMF case of WDLF plotted together with two sets of observational results from \citet{Leggett_1998, Harris_2006} (comparison plot in \citet{Garc_a_Berro_2016}). We normalized the observational data to average total number of WDs in our model ($1.82 \times 10^{10}$). We can see that the general shape of the observationally obtained LF is in agreement with our results. The maximum possible discrepancy in the faint end drop is estimated to be around 0.6. The uncertainties on horizontal axes of both observational results are not available in the source literature. The vertical uncertainties are comparatively large, especially at the faint end. The magnitude uncertainties of the individual WDs in the observations are estimated to reach 0.6, which is comparable with discrepancies with our model \citep{Leggett_1998, Harris_2006}. 

\subsection{Population III white dwarfs signature}
\label{subsec:signature}

\begin{figure}
	\includegraphics[width=\columnwidth]{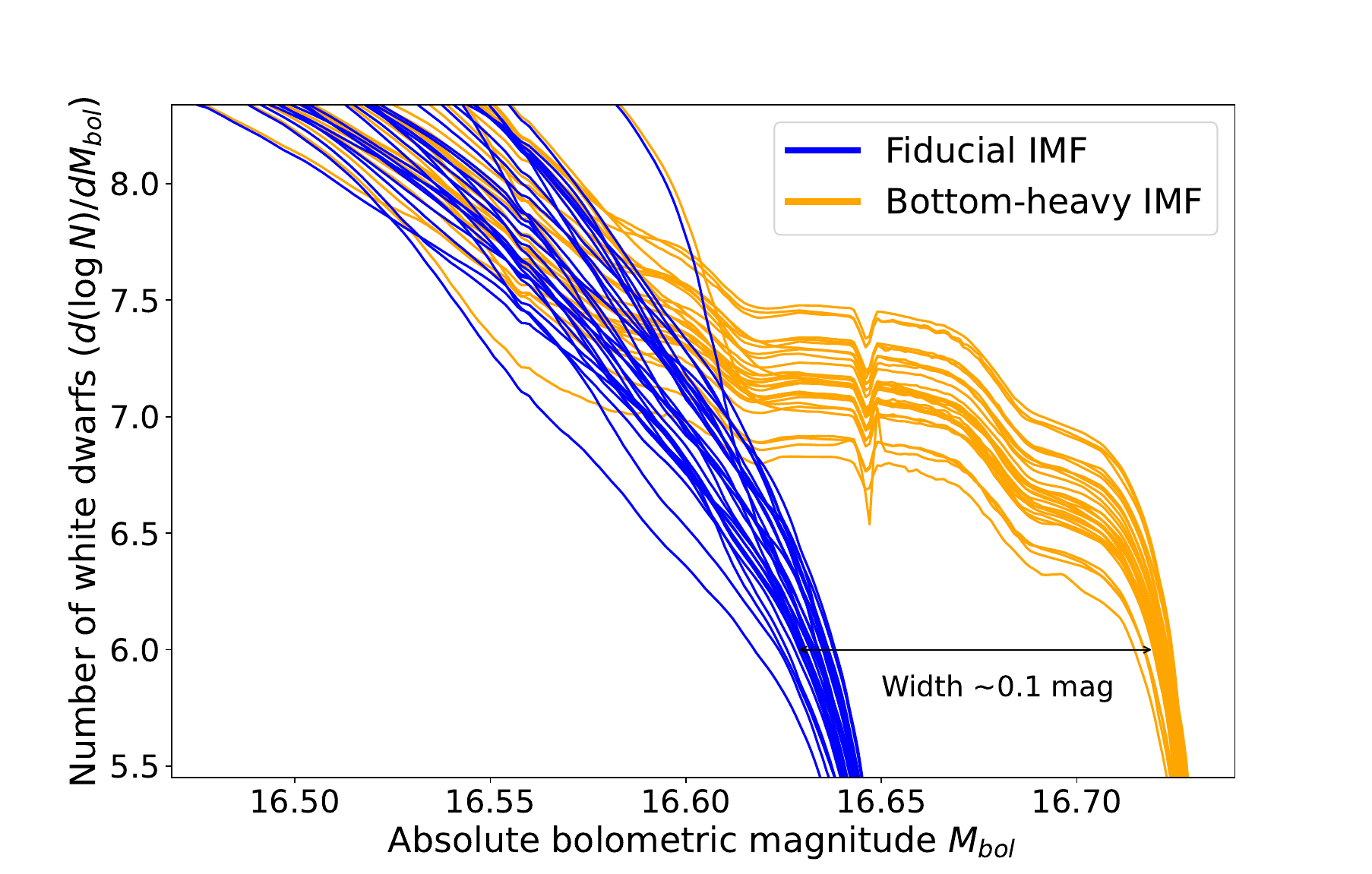}
    \caption{Faint end of WDLF. The blue lines represent WDLF for fiducial Pop~III IMF case while the orange lines represent the case of bottom-heavy IMF. The arrow at $\mathrm{d} \log N/\mathrm{d}M(M) = 6$ shows the difference between two models, approximately 0.1. The small dip at  $M_{\mathrm{abs}} \approx 16.65$ is a resolution effect that is difficult to remove and does not affect the results significantly. It does not convey any physics.}
    \label{fig:II_III_all_zoom}
\end{figure}

\begin{figure}
	\includegraphics[width=\columnwidth]{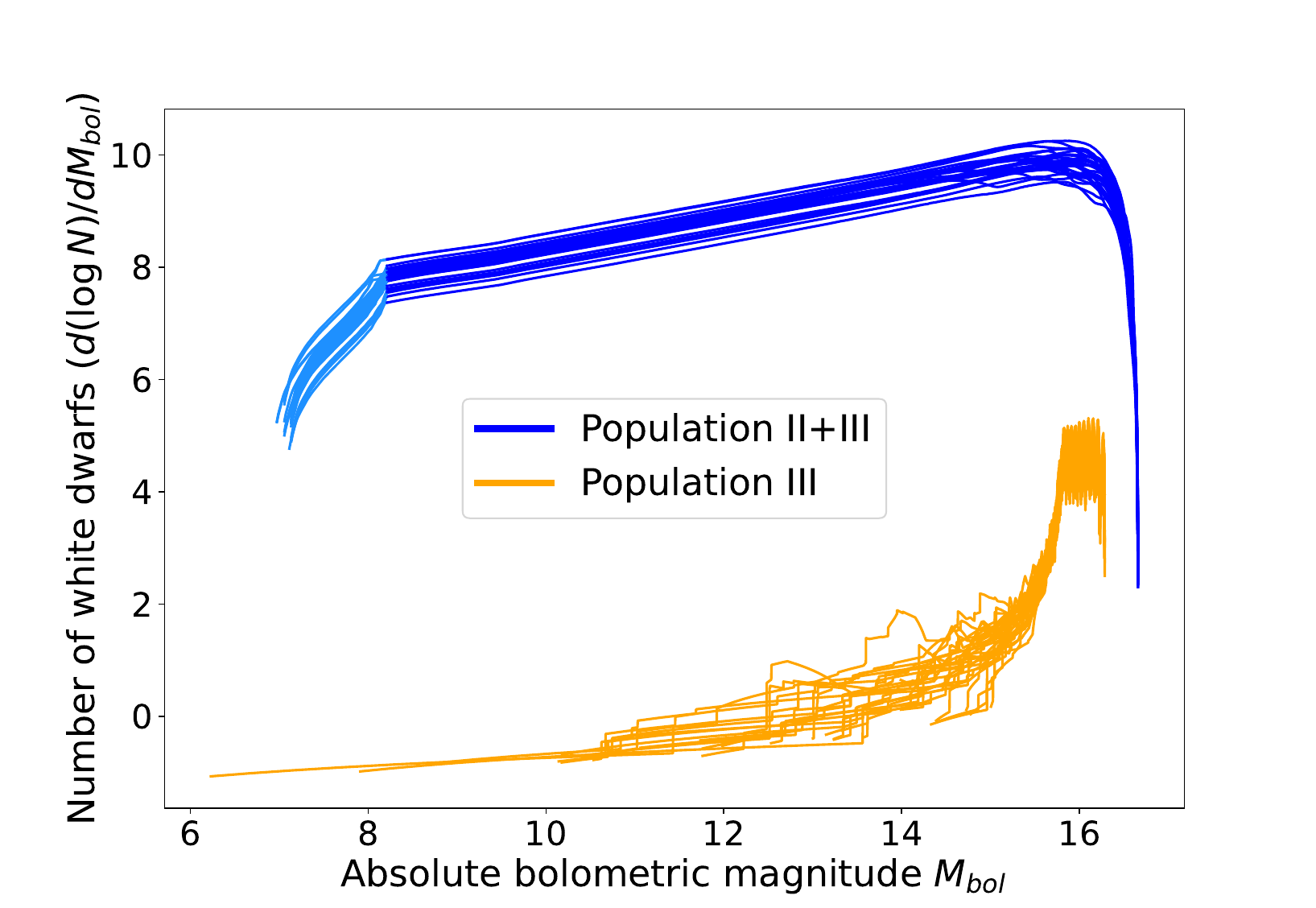}
	\includegraphics[width=\columnwidth]{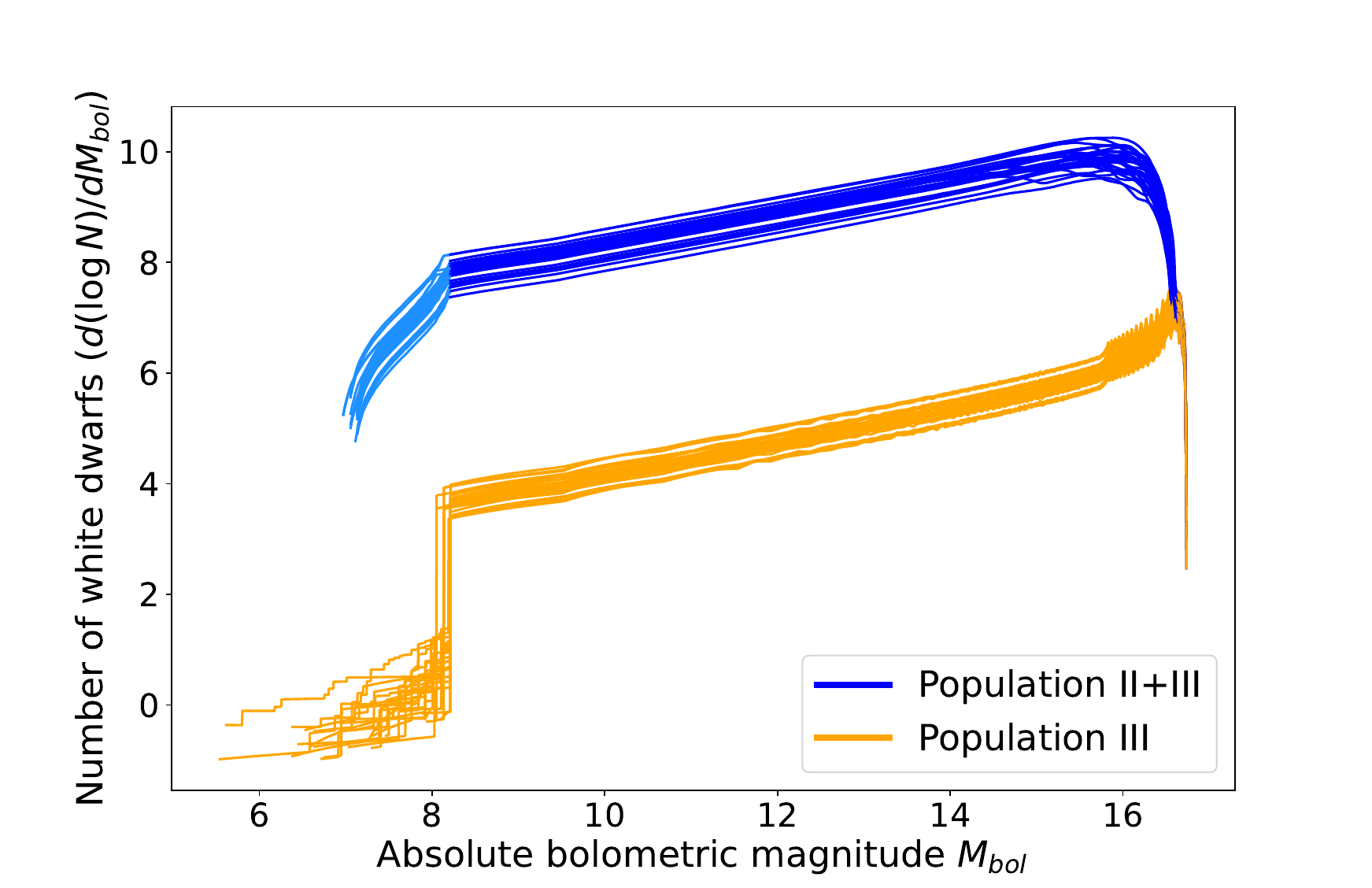}
    \caption{Obtained WDLF divided with respect to progenitor star population. The results are presented separately for (a) fiducial and (b) bottom-heavy cases of Pop~III star IMF. On both plots the blue lines represent the data for all WDs while the yellow lines include only the data for WDs descending from Pop~III stars. By comparing the figures it becomes clear that while for the fiducial IMF case the influence of Pop~III star-descending WDs on the overall IMFR is little, for the bottom-heavy stars it is important, especially in the faint end. The parts of the plots presented in light blue are a linear approximation which influenced its shape}
    \label{fig:II_III_sep}
\end{figure}

Looking at the faint end of LFs, we can notice a hump for the case of bottom-heavy IMF, which is not visible for fiducial IMF. Fig.~\ref{fig:II_III_all_zoom} presents a zoomed-in faint end part for bottom-heavy IMF case together with fiducial IMF case. This hump of width of around $0.1$ is a signature of Pop~III WDs. The part of the hump at $M_{\mathrm{abs}} \approx 16.65$ is simply a resolution effect that is difficult to remove and does not affect the results significantly. It does not convey any physics. Fig.~\ref{fig:II_III_sep} presents the total LFs together with Pop~III-only WD LFs for fiducial and bottom-heavy IMF. The bottom panel shows that the discussed signature is indeed a result of adding the Pop~III WDs to the total LF. Fig.~\ref{fig:III_only} in Appendix presents LFs obtained for Pop~III WDs only. 

\section{Discussion}

\subsection{Mestel's cooling model}
\label{subsec:mesamestel}

As Mestel's model is based on simplified assumptions, we compared its cooling curve to more advanced MESA \citep{Paxton_2010} simulation results. The MESA model that we used for comparison takes into account the element diffusion in a white dwarf, but it does not apply crystallization and phase separation. We first compared the results for a case of a 0.6\Msun~WD. We found that while there are some qualitative differences, both functions are widely consistent for WD ages not exceeding lifetime of the Universe (see Fig.~\ref{fig:mesamestelcomp} in Appendix). The comparison was done for 0.6\Msun~because it is a dominant mass of white dwarf population; most of white dwarfs masses are close to this number. We also performed the same comparison considering the white dwarfs that possibly contribute to the Pop III hump in WDLF. As can be seen on Fig.~\ref{fig:m_comparison_ext} in Appendix, WDs coming from progenitors of masses between 1.25\Msun~and 2.25\Msun~are the ones contributing the most to the hump. It would correspond to WD masses of approximately 0.53\Msun~to 0.64\Msun~according to linear IFMR which we use. The MESA simulation was performed, starting with a 2.0\Msun~progenitor, which produced a 0.50\Msun~white dwarf. The result of cooling simulation is presented on Fig.~\ref{fig:mesamestelcomp}. Again, for 0.5\Msun~white dwarf the MESA and Mestel cooling functions were generally consistent for WD ages not exceeding lifetime of the Universe.

\subsection{Pop III IFMR}
\label{subsec:pop3ifmr}

\begin{figure}
	\includegraphics[width=\columnwidth]{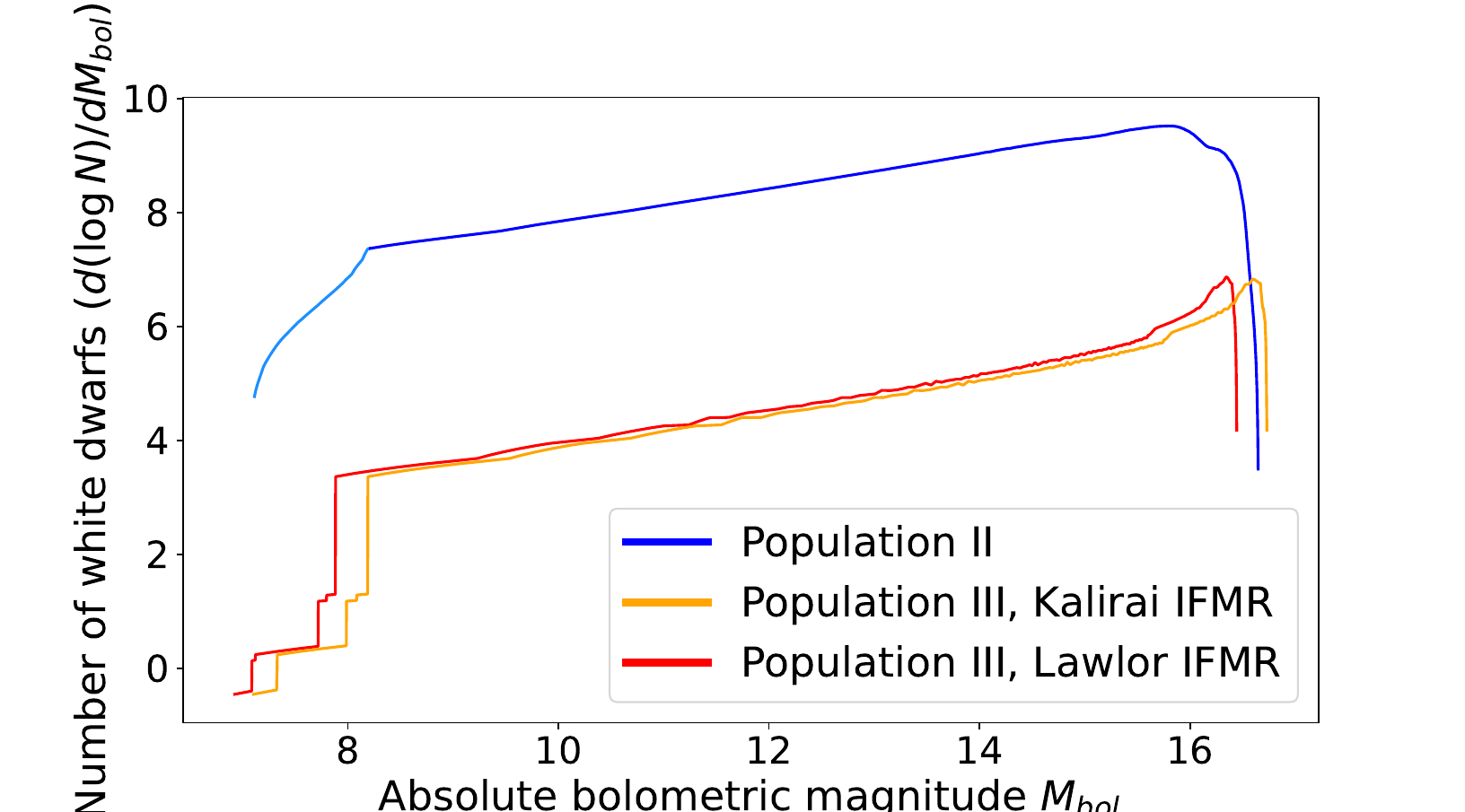}
        \includegraphics[width=\columnwidth]{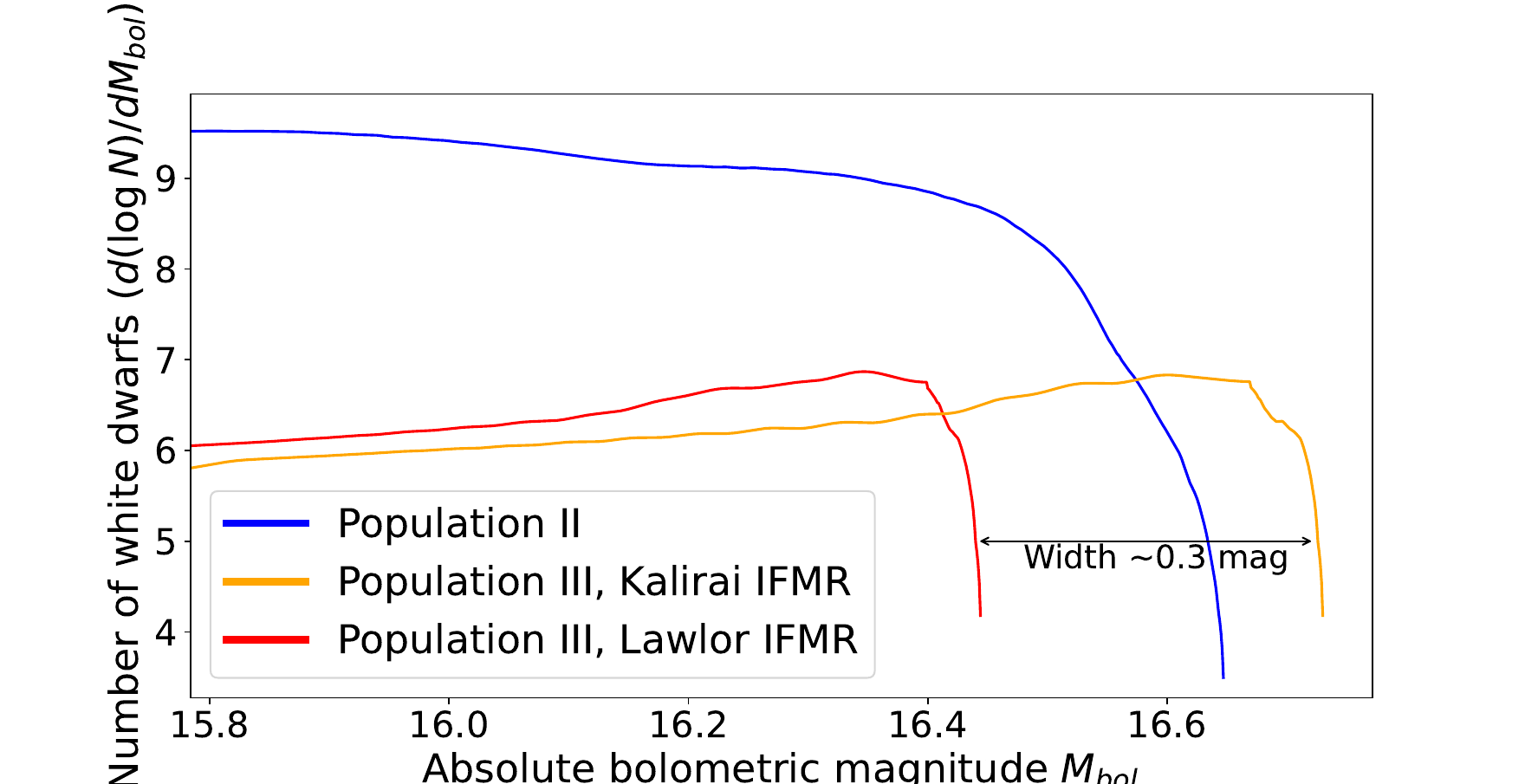}
    \caption{Comparison of a) WDLF b) zoomed faint end of WDLF Kalirai IFMR and Lawlor IFMR. The yellow line represents the WDLF obtained for normally used Kalirai IFMR (slope ($a=0.109$)) for both Pop II and Pop III stars, the red line represents the one obtained for the Lawlor IFMR for Pop III stars, approximated with two linear functions: $M_{\mathrm{WD}}=1.018M_{0} - 0.382$ for $M_{0} < 1$ and $M_{\mathrm{WD}}=0.115M_{0} + 0.556$ for $M_{0} >= 1$ \citep{lawlor13}. The faint ends of the Pop III LFs were smoothed for better visibility. For Lawlor IFMR the faint end of Pop III WDLF does not reach the faint end of Pop II WDLF and therefore the hump becomes invisible.}
    \label{fig:LawlorIFMRcomp_ext}
\end{figure}

The use of Kalirai linear IFMR for Pop III stars resulting from the lack of works on Pop III star IFMR is one of the most serious caveats of this work. Indeed, the information about Pop III IFMR is very scarce and all of the available results are, for obvious reasons, theoretical. A recent work by T. Lawlor and J. MacDonald presents a theoretical IFMR for Pop III stars in a mass range of 0.8-3.0\Msun~\citep{lawlor13}. Although the Authors acknowledge significant uncertainties in their model and the mass range is not sufficient for our work, we decided to examine how would such a Pop III IFMR affect the predicted hump. The IFMR for Pop III stars from \citep{lawlor13} was approximated with two linear functions: $M_{\mathrm{WD}}=1.018M_{0} - 0.382$ for $M_{0} < 1 \Msun$ and $M_{\mathrm{WD}}=0.115M_{0} + 0.556$ for $M_{0} >= 1 \Msun$ and the WDLF calculations were repeated. Note that the second linear function was out of necessity extended for progenitors of $M_{0} > 3 \Msun$. The results are presented on Fig.~\ref{fig:LawlorIFMRcomp_ext}. We can clearly see that for Lawlor IFMR the faint end of Pop III WDLF does not reach as low magnitudes as the faint end of Pop II WDLF and therefore the hump would not be visible. The shift of the Pop III WDLF for Lawlor IFMR towards the brighter magnitudes can be easily understood considering the fact that for the most of the progenitor masses the Lawlor IFMR predicts higher WD masses than the Kalirai IFMR. For Mestel's model (see equation~(\ref{eq:mestel}) in Section~\ref{subsubsec:wdcoolingfun}) it simply means that at a given time the luminosity of the corresponding white dwarfs will be higher in Lawlor IFMR case than in Kalirai IFMR case. In particular, it is true for all the progenitors of masses higher than 1 \Msun, so for all the white dwarfs that would potentially contribute to the Pop III hump.

\subsection{Possibility of Pop III signature detection}
\label{subsec:humpdetection}

Potential observation of the Pop~III signature in WDLF introduced in \ref{subsec:lumfun} would become the first ever evidence of low-mass Pop~III stars. It would also let us constrain the Pop~III IMF through the number of low-mass Pop~III stars formed in MW. Contrastingly, if the WDLF obtained with precise and complete observation(s) would not show such signatures, it would also put a constraint on the number of low mass Pop~III stars. 

Assuming that the real IMF is indeed closer to the bottom-heavy case, let us  discuss the possibilities of observational detection of Pop~III WD signature.
The values related to the position of the LF hump are presented together in Table~\ref{tab:summary}. We measure all the x-axis intervals (magnitude) at the same $y$-value of 6. The cosmic variance itself does not affect the possibility of signature detection, but marginally changes its expected position (while significantly affecting the absolute number of WDs). This change is estimated to not exceed 0.01. To precisely predict the shape and position of the signature, we would like the uncertainties connected to IFMR and cooling model to be as low as possible, ideally to sum up to value significantly lower than the width of the signature (0.1). The main restriction on the detection possibility is our observational potential -- one can see that current observations, while heavily uncertain, differ from our model in faint-end drop-off position as much as 0.6. Again, we would need the observational data with absolute magnitude precision better than 0.1. Currently, the biggest hope for significantly better quality data are the recent and upcoming releases of Gaia mission -- current releases can measure the faint stars magnitudes with great precision of around 0.001 and the biggest error in absolute magnitudes comes from parallax. Especially for relatively close stars and low extinction cases, the uncertainty might be way lower than predicted Pop~III signature width \citep{GAIAHR}. However, the downside of the Gaia mission is its relatively low limiting magnitude of 20.7. The faintest (Pop~III) WDs in our model have $M_{\mathrm{abs}} \approx 17$ and so we can assume that the range in which a WD survey is complete is the range in which objects of $M_{\mathrm{abs}} = 17$ are within the limiting magnitude of the survey. It means that Gaia WD data can be complete only in the vicinity of 55 pc. Assuming that the total number of WDs is proportional to the volume of the sample, in that close vicinity less than 3000 WDs are expected to reside. Given the shares of Pop~III WDs in overall population in our model, we could then expect no Pop~III WDs in fiducial case and around one Pop~III WD in bottom-heavy case in this vicinity. The more promising are Euclid telescope (recently launched) and the Vera C. Rubin Observatory (set to be launched in early 2025). Euclid has a visual limiting magnitude of 26.2 \citep{euclid} while Rubin is expected to reach as deep as 27 \citep{rubin}. This would mean that the WD survey could be complete for the vicinity of 690 pc and 1000 pc respectively, giving us the probe of millions of WDs. Then, while in bottom-heavy case we could expect thousands of Pop~III WD detections, even in the fiducial case a few to a few dozens WDs could be detected. Data from these surveys could possibly answer the question of the presence of the Pop~III WD hump and potentially help us detect the first Pop~III WDs. 

Gaia and Rubin missions provide the multicolor photometry of the target objects. It can be a useful tool to identifying white dwarfs and therefore helpful in obtaining an accurate WDLF. Using multicolor photometry can also improve completeness of a WDLF as the populations are viewed through different color bands. Moreover, observational band-limited WDLF can be created. This approach can be used in creating more selective LFs, for example for only cool WDs, which are contributing to the potential hump. It is, however, difficult to predict whether the qualitative shape of the WDLF faint end would differ significantly in such approach because little is known about the differences of Pop II and Pop III WD spectra.

\begin{figure}
	\includegraphics[width=\columnwidth]{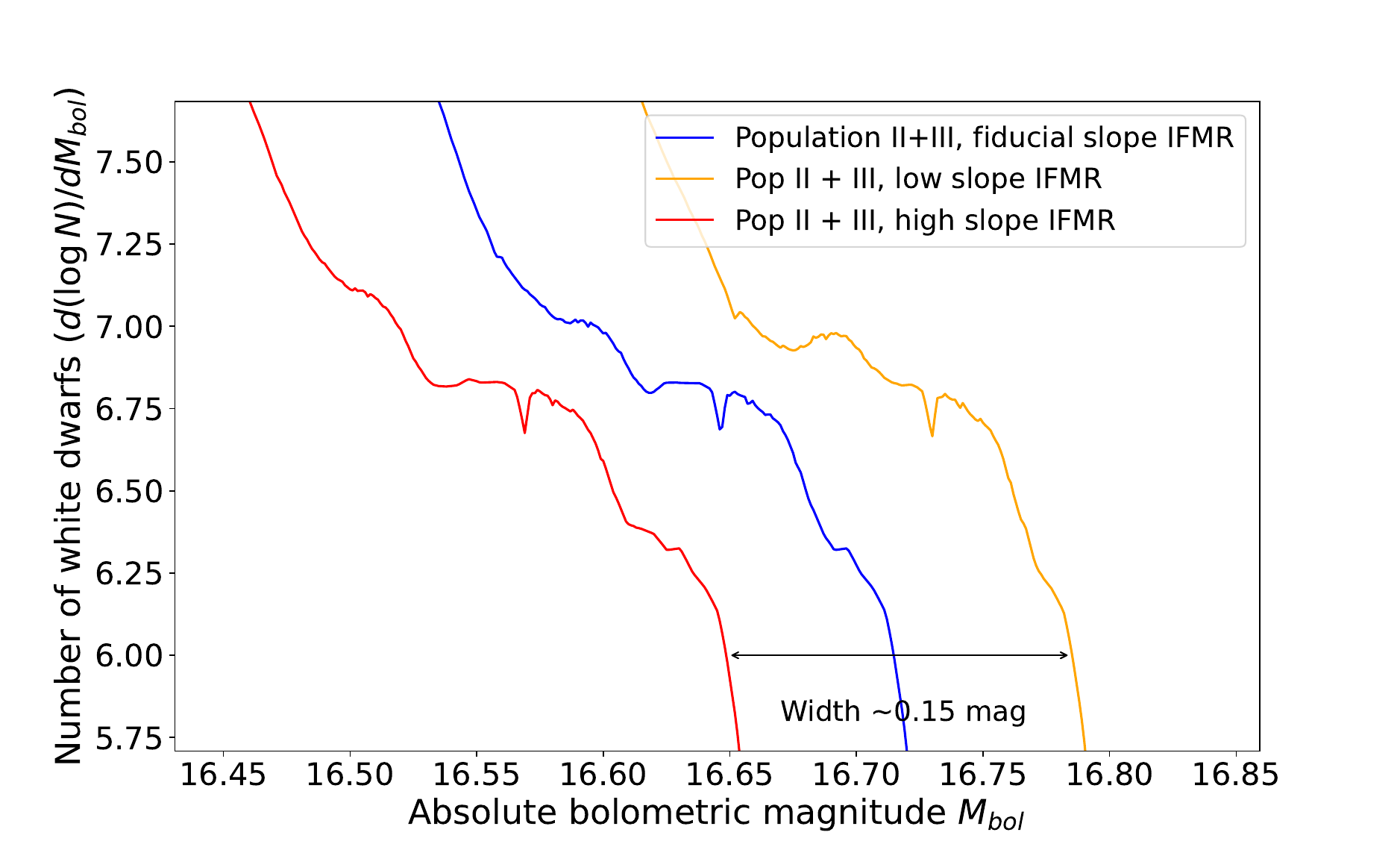}
    \caption{Comparison of faint end of WDLF with extreme values of slope in linear IFMR. The blue line represents the WDLF obtained for normally used IFMR (slope ($a=0.109$)), the yellow line represents the one obtained for the lowest slope within the 3x uncertainty range ($a=0.88$) and red line represents results for the highest slope ($a=0.130$).}
    \label{fig:extremeIFMRcomp_ext}
\end{figure}

\begin{figure}
	\includegraphics[width=\columnwidth]{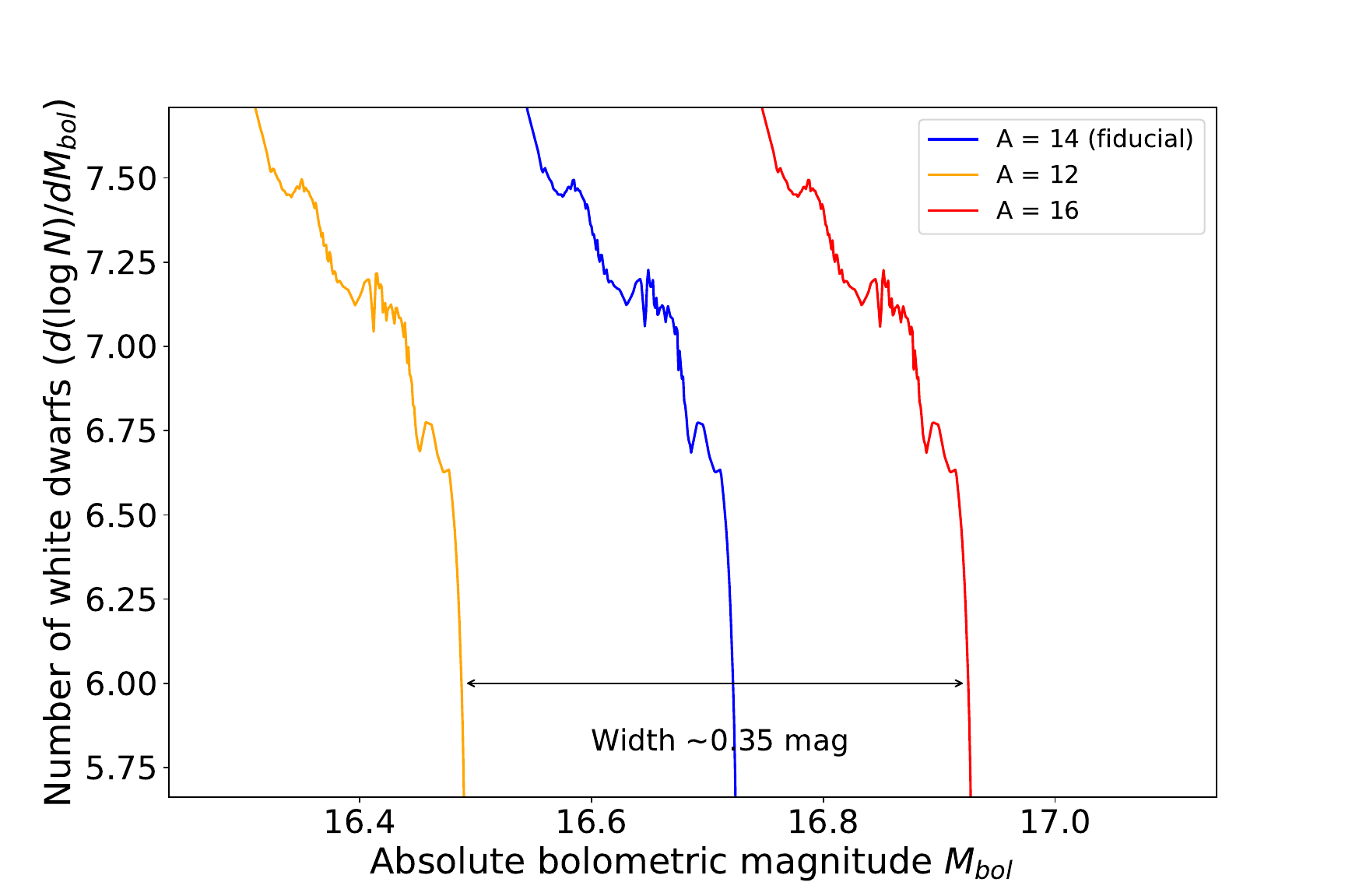}
    \caption{omparison of faint end of WDLF with extreme values of $A$ in Mestel's model equation. The blue line represents the WDLF obtained for the value of $A$ between carbon and oxygen ($A=14$) while the yellow and red line represents the extreme carbon-only ($A=12$) and oxygen-only ($A=16$) cases respectively.}
    \label{fig:extremeAcomp_ext}
\end{figure}

\begin{table}
    \centering
    \caption{Summary table comparing the main uncertainties and features that we can track on the horizontal axis of obtained WDLF. The smallest value named "Cosmic variance" represent the x-axis differences between the thirty MW realizations in our model.}
    \label{tab:summary}
    \begin{tabular}{cc}
%        \hline
%		Value &  [mag] \\
        \hline
        Gaia uncertainty & 0.05 mag \\
        Discrepancy from observations & 0.6 mag \\
        WD core chemical composition uncertainty & 0.35 mag \\
        IFMR uncertainty & 0.15 mag \\
        Cosmic variance & 0.01 mag \\
        Pop III signature & 0.1 mag \\
        \hline
    \end{tabular}
\end{table}

\section{Conclusions}

Applying a simple WD cooling model to the WD populations simulated with \textsc{a-sloth}, we have derived WDLFs that are consistent with other models and observations. Our model also predicts a previously unknown LF feature -- a hump being a signature of Pop~III stars. Further studies are clearly needed from two different perspectives. Firstly, more advanced models should be developed by employing, for instance, an updated WD cooling model, to predict accurately the exact shape of the LF. Such models should also incorporate the detailed features of Initial-Final Mass Relations, including its dependence on progenitor star metallicity and the non-monotonic parts. We note that this might not be straightforward without the Initial-Final Mass Relation itself being greatly improved. In order to compare with observational data in a more careful manner, disk and halo WDs should be distinguished in the model. At the same time, as \textsc{a-sloth} does not consider the formation of binary systems, it would also be necessary to include the contribution of binary WDs on the LF (see Section~\ref{sec:introduction}). 

Future observational data, including data from latest GAIA releases as well as data from Euclid and LSST, can be used to derive more accurate WDLFs of quality greatly exceeding the currently available one. The faint end of the LF should be carefully analyzed in search of any signature of Pop~III stars. In the future, confirming or excluding the faint-end hump of the WDLF can possibly place constraints on the formation efficiency and the mass distribution of Pop~III stars. 

\section*{Acknowledgements}
We acknowledge funding from JSPS KAKENHI Grant Number 20K14464. TH is also employed at the German Environment Agency.

%%%%%%%%%%%%%%%%%%%%%%%%%%%%%%%%%%%%%%%%%%%%%%%%%%
\section*{Data Availability}
 
Data is available upon request.

%%%%%%%%%%%%%%%%%%%% REFERENCES %%%%%%%%%%%%%%%%%%

\bibliographystyle{mnras}
\bibliography{biblio}

%%%%%%%%%%%%%%%%%%%%%%%%%%%%%%%%%%%%%%%%%%%%%%%%%%

%%%%%%%%%%%%%%%%% APPENDICES %%%%%%%%%%%%%%%%%%%%%

\appendix

\section{Additionally produced plots}

\begin{figure}
	\includegraphics[width=\columnwidth]{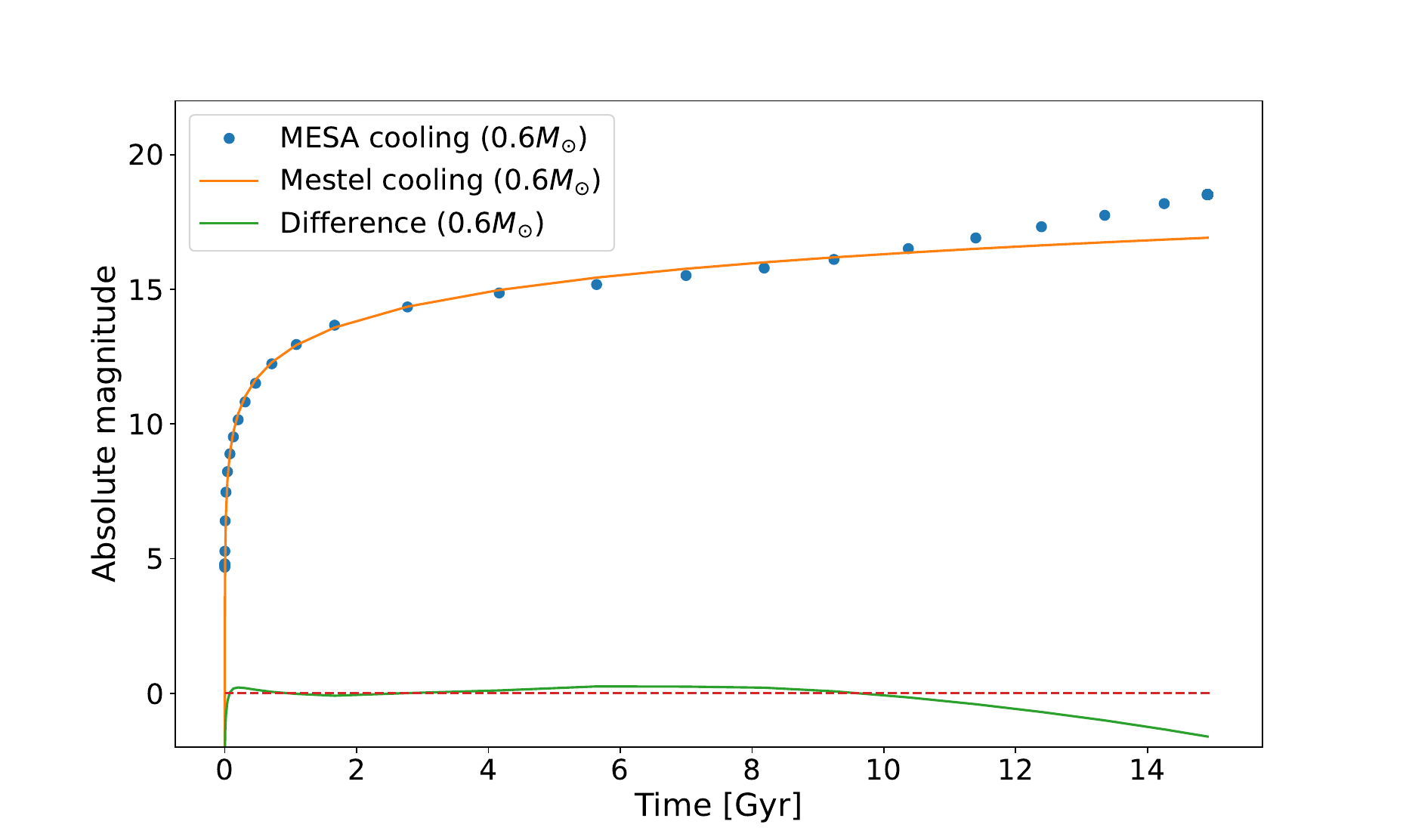}
        \includegraphics[width=\columnwidth]{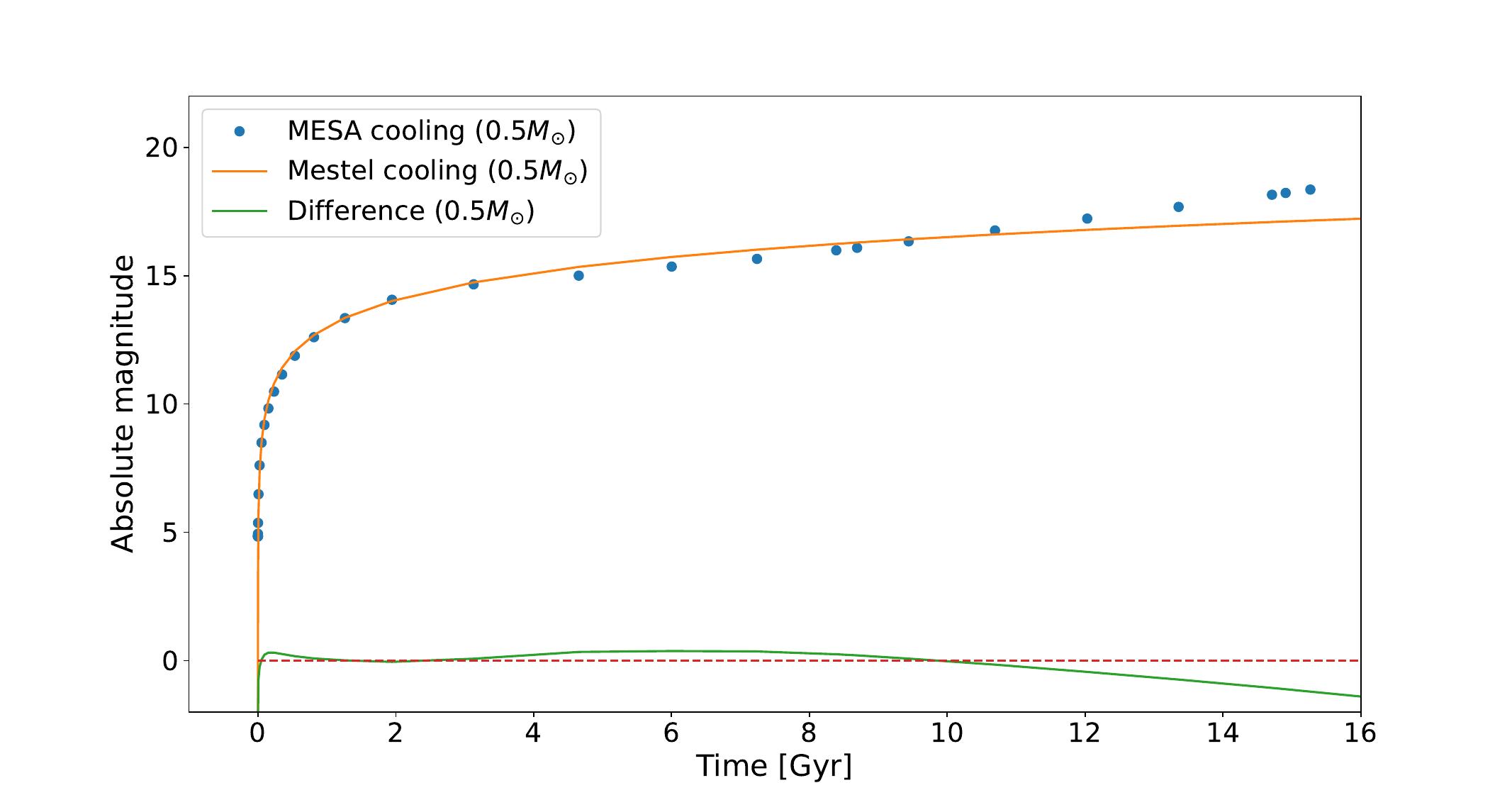}
    \label{fig:mesamestelcomp}
    \caption{The cooling curve of a) 0.6\Msun~WD b) 0.5\Msun~WD obtained with Mestel's model (orange line) compared to MESA simulation (blue dots). The differences (green line) do not exceed 1 for the most of the cooling time.}
\end{figure}

\begin{figure}
    \centering
	\includegraphics[width=\columnwidth]{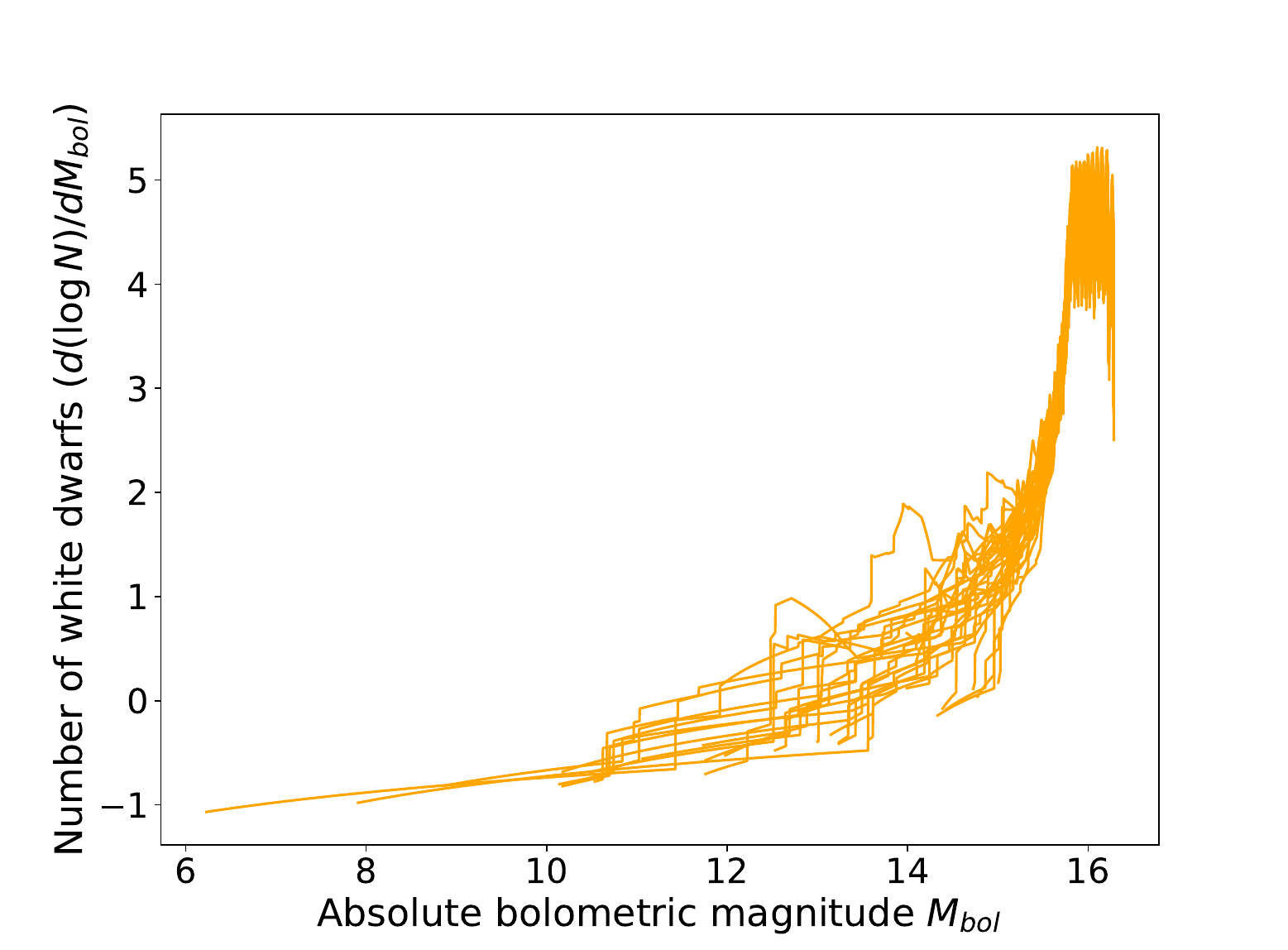}
	\includegraphics[width=\columnwidth]{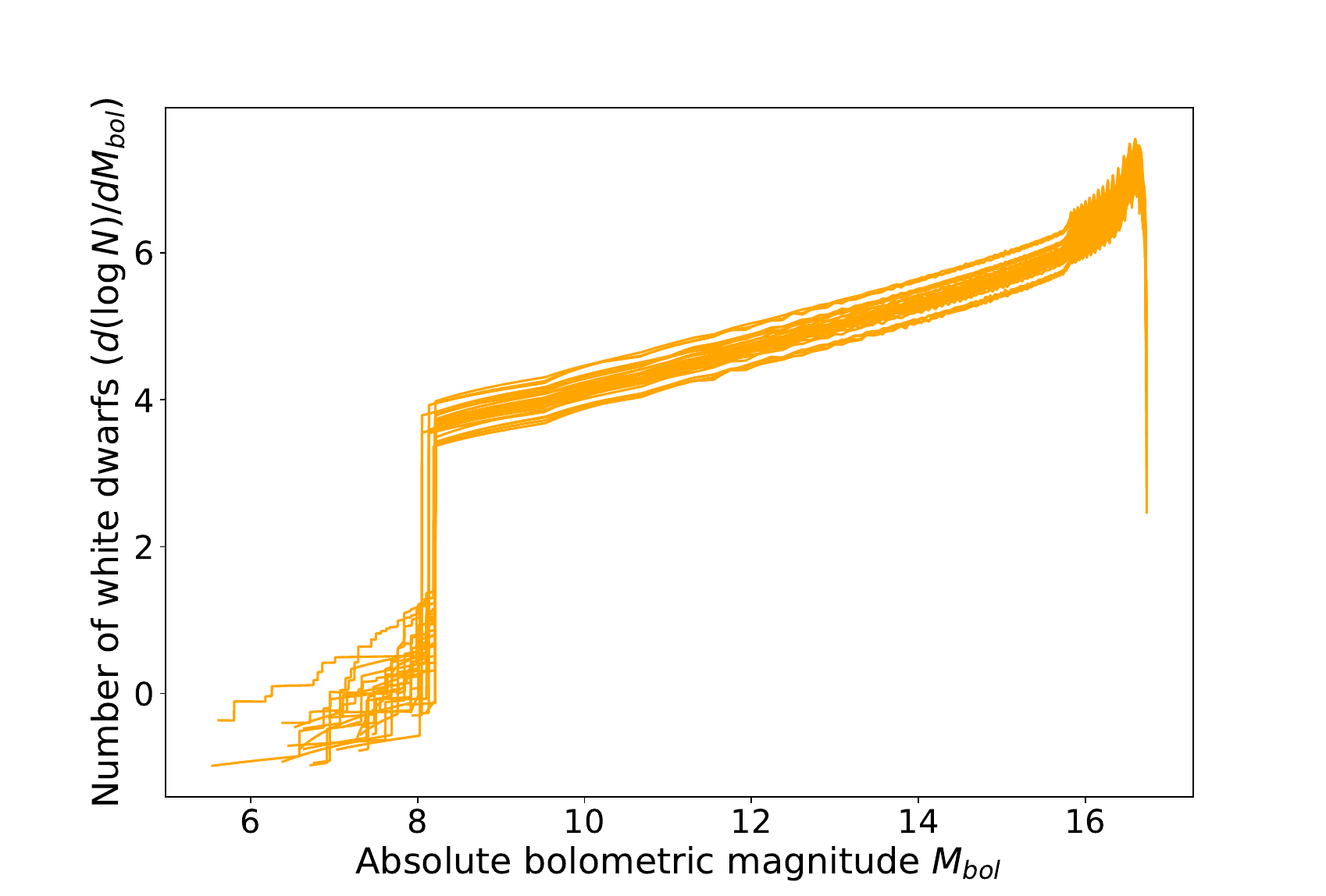}
    \label{fig:III_only}
    \caption{Obtained WDLF limited to Pop~III star descendants for (a) fiducial and (b) bottom-heavy cases of Pop~III star IMF. On both plots different orange lines correspond to different MW-like galaxy realizations.}
\end{figure}

\begin{figure}
	\includegraphics[width=\columnwidth]{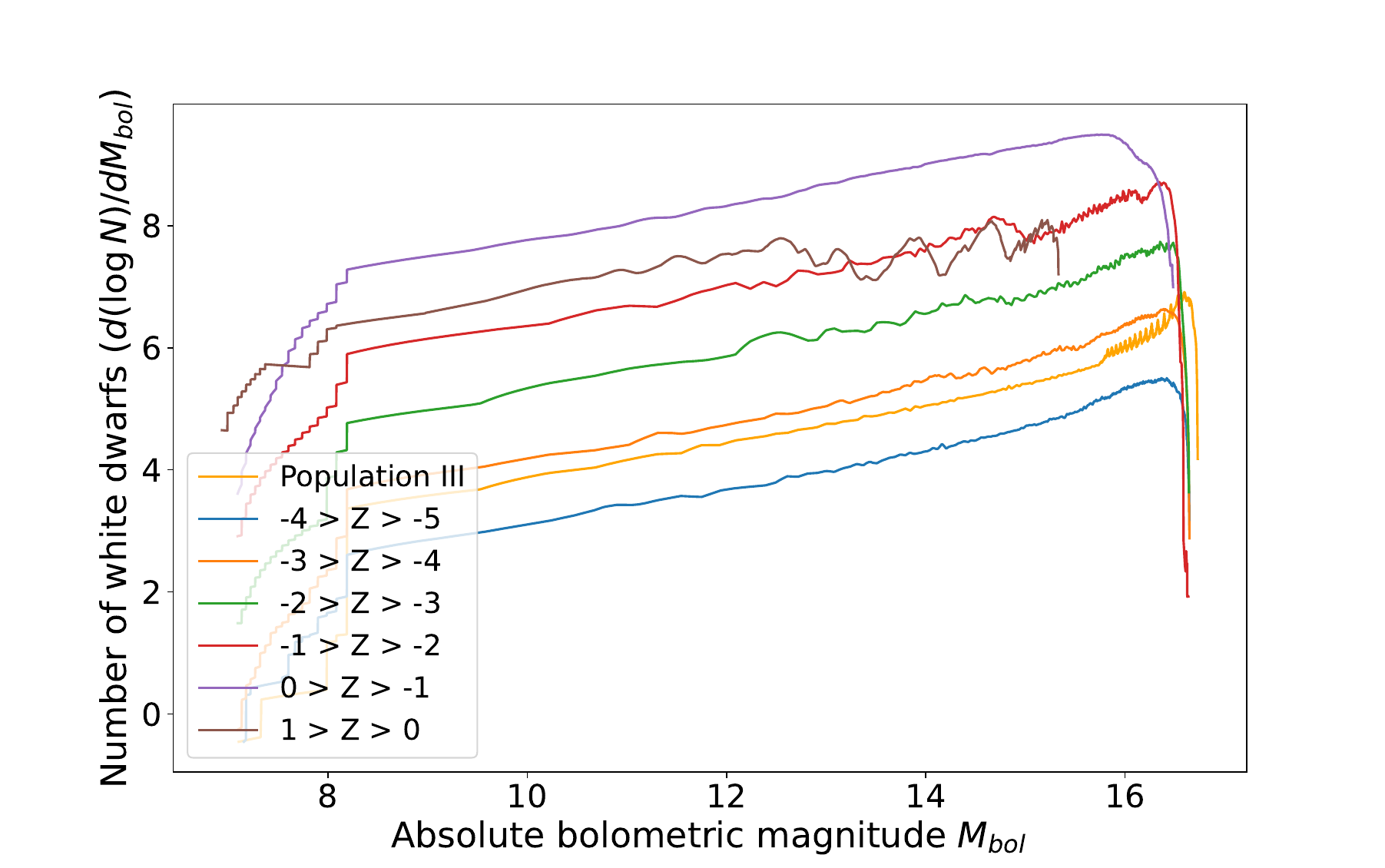}
    \caption{WDLF divided with respect to progenitor star metallicity. The plot was created for fiducial Pop~III IMF case and only one MW-like galaxy realization was chosen to ensure clarity. Lines of different colours represent different progenitor star metallicity bins with Pop~III star descendants marked separately with yellow line.}
    \label{fig:z_comparison_ext}
\end{figure}

\begin{figure}
	\includegraphics[width=\columnwidth]{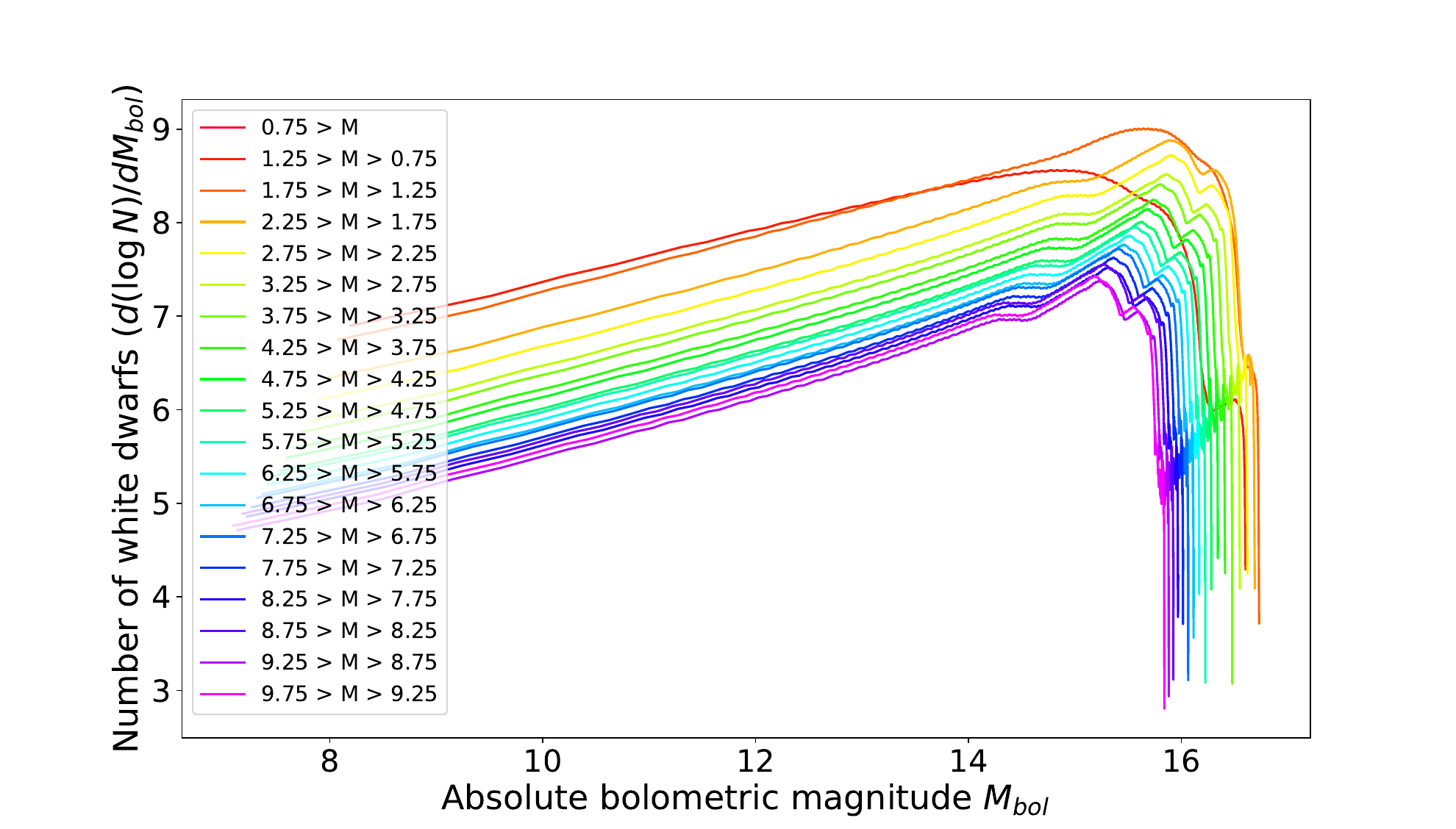}
    \caption{WDLF divided with respect to progenitor star mass. The plot was created for fiducial Pop~III IMF case and only one MW-like galaxy realization was chosen to ensure clarity. Lines of different colours represent different progenitor star mass bins.}
    \label{fig:m_comparison_ext}
\end{figure}

%%%%%%%%%%%%%%%%%%%%%%%%%%%%%%%%%%%%%%%%%%%%%%%%%%

% Don't change these lines
\bsp	% typesetting comment
\label{lastpage}
\end{document}